\documentclass[conference]{IEEEtran}
\IEEEoverridecommandlockouts
\usepackage{cite}
\usepackage{amsmath,amssymb,amsfonts}
\usepackage{algorithmic}
\usepackage{graphicx}
\usepackage{textcomp}
\usepackage{xcolor}

\def\BibTeX{{\rm B\kern-.05em{\sc i\kern-.025em b}\kern-.08em
    T\kern-.1667em\lower.7ex\hbox{E}\kern-.125emX}}
\begin{document}

\title{A SSIM Guided cGAN Architecture For Clinically Driven Generative Image Synthesis of Multiplexed Spatial Proteomics Channels\\

\thanks{This work was supported by a University of Texas System Rising STARs Award (J.M.L) and the CPRIT First Time Faculty Award (J.M.L).}
}

\author{\IEEEauthorblockN{Jillur Rahman Saurav\IEEEauthorrefmark{1}\IEEEauthorrefmark{2}, Mohammad Sadegh Nasr\IEEEauthorrefmark{1}\IEEEauthorrefmark{2}, Helen H Shang\IEEEauthorrefmark{5}, Paul Koomey\IEEEauthorrefmark{1}\IEEEauthorrefmark{2}, \\ Michael Robben\IEEEauthorrefmark{1}\IEEEauthorrefmark{2}, Manfred Huber\IEEEauthorrefmark{1}, Jon Weidanz\IEEEauthorrefmark{2}\IEEEauthorrefmark{3}\IEEEauthorrefmark{4}, Br\'{i}d Ryan\IEEEauthorrefmark{6}, \\  Eytan Ruppin\IEEEauthorrefmark{7}, Peng Jiang\IEEEauthorrefmark{7}, Jacob M. Luber\IEEEauthorrefmark{1}\IEEEauthorrefmark{2}\IEEEauthorrefmark{3}}
\IEEEauthorblockA{\IEEEauthorrefmark{1}Department of Computer Science, University of Texas at Arlington}
\IEEEauthorblockA{\IEEEauthorrefmark{2} Multi-Interprofessional Center for Health Informatics, University of Texas at Arlington }
\IEEEauthorblockA{\IEEEauthorrefmark{3}Department of Bioengineering, University of Texas at Arlington}
\IEEEauthorblockA{\IEEEauthorrefmark{4}Department of Kinesiology, University of Texas at Arlington}
\IEEEauthorblockA{\IEEEauthorrefmark{5}Division of Internal Medicine, Ronald Reagan University of California Los Angeles Medical Center}
\IEEEauthorblockA{\IEEEauthorrefmark{6}Laboratory of Human Carcinogenesis, Center for Cancer Research, National Cancer Institute, National Institutes of Health}
\IEEEauthorblockA{\IEEEauthorrefmark{7}Cancer Data Science Laboratory, Center for Cancer Research, National Cancer Institute, National Institutes of Health
}
\IEEEauthorblockA{Email: {\{mdjillurrahman.saurav, mohammadsadegh.nasr}\}@uta.edu, hshang@mednet.ucla.edu, \{paul.koomey, michael.robben\}@uta.edu, \\ huber@cse.uta.edu, weidanz@uta.edu,\{Brid\_Ryan, eytan.ruppin, peng.jiang\}@nih.gov, jacob.luber@uta.edu
}
}

\maketitle

\begin{abstract}
Histopathological work in clinical labs often relies on immunostaining of proteins, which can be time-consuming and costly. Multiplexed spatial proteomics imaging can increase interpretive power, but current methods cannot cost-effectively sample the entire proteomic retinue important to diagnostic medicine or drug development. To address this challenge, we developed a conditional generative adversarial network (cGAN) that performs image-to-image (i2i) synthesis to generate accurate biomarker channels in multiplexed spatial proteomics images\footnote{https://github.com/aauthors131/multiplexed-image-synthesis}.We approached this problem as missing biomarker expression generation, where we assumed that a given n-channel multiplexed image has p channels (biomarkers) present and q channels (biomarkers) absent, with p+q=n, and we aimed to generate the missing q channels. To improve accuracy, we selected p and q channels based on their structural similarity, as measured by a structural similarity index measure (SSIM). We demonstrated the effectiveness of our approach using spatial proteomic data from the Human BioMolecular Atlas Program (HuBMAP)\footnote{https://portal.hubmapconsortium.org}, which we used to generate spatial representations of missing proteins through a U-Net based image synthesis pipeline. Channels were hierarchically clustered by SSIM to obtain the minimal set needed to recapitulate the underlying biology represented by the spatial landscape of proteins. We also assessed the scalability of our algorithm using regression slope analysis, which showed that it can generate increasing numbers of missing biomarkers in multiplexed spatial proteomics images. Furthermore, we validated our approach by generating a new spatial proteomics data set from human lung adenocarcinoma tissue sections and showed that our model could accurately synthesize the missing channels from this new data set. Overall, our approach provides a cost-effective and time-efficient alternative to traditional immunostaining methods for generating missing biomarker channels, while also increasing the amount of data that can be generated through experiments. This has important implications for the future of medical diagnostics and drug development, and raises important questions about the ethical implications of utilizing data produced by generative image synthesis in the clinical setting.

\end{abstract}


\section{Introduction}
\label{sec:intro}
    % For review, code and data availability information is available at \url{https://github.com/aauthors131/multiplexed-image-synthesis}.\\
    Immunohistochemical (IHC) techniques have long been used in the visualization of tissues for both research and clinical purposes \cite{shi2011antigen}. The shift from chromogenic peroxidase conjugated antibodies to fluorescent reporters was a driving catalyst in the adoption of IHC as a tool for biomarkers of human disease \cite{tada2016potential,tan2020overview}. Multiplexed IHC (mIHC) of CDX2 and SOX2 in histological stains has successfully been used as a marker for primary colorectal cancer \cite{lopes2020digital}. Unfortunately, due to restrictions in fluorescent microscopy, traditional Immunofluorescence (IF) mIHC techniques are limited to 3 protein markers in addition to the nucleic marker molecule 4',6-diamidino-2-phenylindole (DAPI) \cite{stack2014multiplexed}. Flow cytometry has revealed that accurate characterization of certain cell populations requires at least 12-18 unique markers \cite{chattopadhyay2012cytometry}. New techniques in DNA oligo based fluorescent reporters, such as CODEX/Phenocylcer-Fusion, have increased the number of simultaneous markers from 22-36, with each marker saved as a channel on an n-channel image \cite{goltsev2018deep,schurch2020coordinated}.\\
    Computational analysis of IHC images has also progressed significantly since the adoption of machine learning into biological analysis workflows. Canonical methods of cell identification by hand or color segmentation have given way to new edge-based and machine learning guided techniques \cite{di2010automated}. A convolutional neural network using U-Net architecture classified 6 cell types in a colored chromagen IHC stain but required multiple stained slides of different sections to do so \cite{fassler2020deep,ronneberger2015u}. Multiple other studies have applied deep learning segmentation to classification of nuclei, cancer types, and tumor microenvironments \cite{ghoshal2021deephistoclass,bulten2019epithelium,zadeh2021deep}. Current implementations are hampered by low throughput of protein marker visualization or relegated to meta analysis of collective data sets.\\ 
    The problem of multiplexing can be solved by the computational generation of imaging data through Generative Adversarial Networks (GANs). The pix2pix algorithm developed in 2017 \cite{zhu2017unpaired}, has allowed translational generation of images of high photo-realism. GANs have already seen wide adoption in medical imaging, with over 100 papers published so far using GANs to generate MRI, CT, PET, and Tomography data \cite{tschuchnig2020generative,yi2019generative}. \cite{yi2018sharpness}, successfully utilized generated data to de-noise low dose liver CT images, preserving tissue morphology in a clinically relevant way. While such novel uses of GANs may be useful in reducing cost and wait times for life saving medical treatments, there are numerous ethical considerations for using generated data in patient care.\\
    The use of GANs in the generation of protein IHC stains from low multiplexed tissue has not been fully explored. A handful of studies have already demonstrated the ability for GANs to segment cells using multiplexed IHC \cite{gadermayr2018way,gupta2019gan}. From this data, \cite{tschuchnig2020generative} proposed that stain-to-stain translation through GANs would be possible with large, multiplexed tissue stain experiments as training data. \cite{bao2021random} recently developed a random channel GAN algorithm to compute missing data from tissue stained sequentially with 11 different markers but were impeded by the limitations of indirect IF staining. Additionally, recent methods have proposed image to `omics translation algorithms \cite{schapiro2022mcmicro,schapiro2022miti}. 

In this paper, we present a cGAN-based image-to-image translation technique that generates novel protein channels in multiplexed CODEX images. Our method utilizes experimental data from HuBMAP as well as data generated in the wetlab to evaluate its efficacy. This approach significantly extends beyond previous work by \cite{bao2021random}, which only used 11 markers, by utilizing multiplexed data with 29 markers. Our proposed architecture employs an automatic SSIM-based channel selection component that facilitates scaling to higher numbers of channels. We trained models on dozens of markers from the same slide and successfully generated new data in cancerous IHC slides with limited marker multiplexing.
In the following sections, we provide a detailed overview of our proposed model architecture, followed by the implementation and evaluation of generative image synthesis on HUBMAP data. We then discuss the methodology for determining the optimal scaling of the model and selecting training channels using SSIM. Furthermore, we present the clinical validation of our model and showcase clinically relevant generative images. Finally, we provide a discussion of our results and conclusions drawn from our study.
    
    % To our knowledge, we have developed the first cGAN based image to image translation approach to generate novel protein channels in multiplexed CODEX images using experimental data from HuBMAP as well as data that we experimentally generated in the wetlab to evaluate our method; this approach for the first time extends significantly beyond the 11 markers used in previous work by \cite{bao2021random} by utilizing multiplexed data with 29 markers, and can potentially be scaled to hundreds of proteomic markers through the use of the SSIM guided channel selection component of the proposed architecture. Our model uses automatic SSIM-based channel selection to facilitate scaling to higher numbers of channels and obtain better results for individual channels here during generative synthesis. This allowed us to train models on dozens of markers from the same slide, and select markers to generate new data in cancerous IHC slides with limited marker multiplexing. 
    
\section{Overview of Model Architecture}

     \begin{figure*}[ht!]
      \centering
      \includegraphics[scale=0.8]{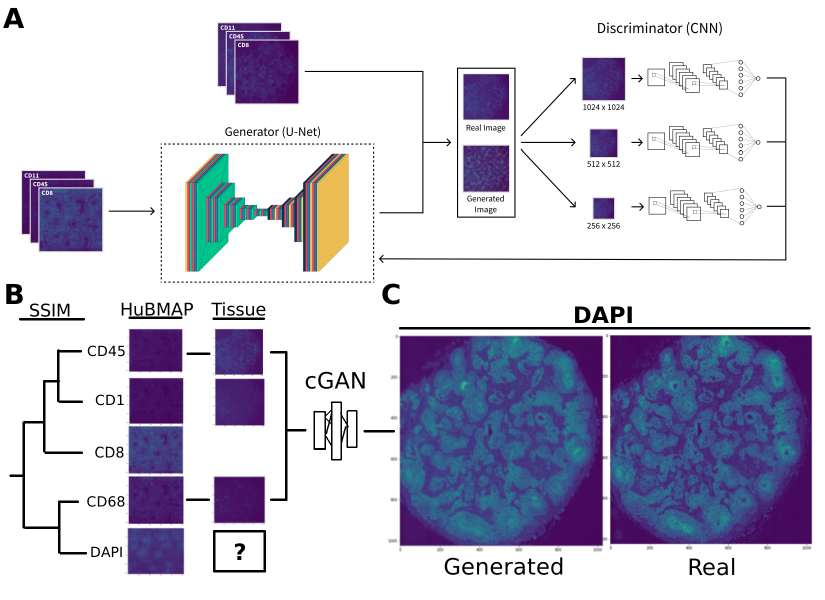}
      \caption{Overview of model architecture showing how DAPI is generated. (A) A cGAN model was used to predict missing channels from multiplexed spatioproteomic data generated on the CODEX/phenocycler platform. Training data sets were divided into training and validation data sets based on random sampling. Iterative models were developed off inclusion of multiple marker channels. U-Net is illustrated using \texttt{visualkeras} package (\cite{Gavrikov2020VisualKeras}) (B) Heuristic clustering by SSIM removed presence bias of marker channels in test image data sets. (C) cGAN accurately predicted DAPI channels in HuBMAP data sets. Generated images were single channeled falsely colored in figure.}
      \label{fig:sum}
    \end{figure*}
    
    % In this work, trained a U-NET based image synthesis pipeline on 8 NVIDIA DGX A100 high memory GPUs to generate photo accurate representations of held out non-cancerous channels experimentally generated with the Akoya Biosciences PhenoCycler System. 
    % through the HUBMAP consortium \cite{hubmap2019human}. 
    Figure \ref{fig:sum}A shows an overview of the U-NET based cGAN network architecture consisting of a generator and discriminator network that was derived from the pix2pixHD architecture \cite{https://doi.org/10.48550/arxiv.1711.11585}. We have used the \cite{jeong2020solar} implementation as the baseline for our network. Rather than random noise like in a GAN network, our cGAN takes in conditional channels as input and tries to predict CODEX/Phenocycler-Fusion channels that were either held out during training or not experimentally collected during inference. Our model improves image synthesis iteratively by the loss function first introduced in \cite{isola2017image} where LGAN (Equation ~\ref{eq:gan_loss}) and L1 (Equation ~\ref{eq:l1_loss}) are used to derive our objective function (Equation ~\ref{eq:objective}), and Generator (G) tries to reduce the objective while Discriminator (D) attempts to achieve greater. Here x,y are the input image and target image.
       
    \begin{equation}
    \begin{split}
        \label{eq:gan_loss}
        \mathcal{L}_{GAN}(G,D) = \mathbb{E}_{y}\left[ \log D(y) \right] \\ + \mathbb{E}_{x}\left[ \log \left( 1 - D(G(x)) \right) \right]
    \end{split}
    \end{equation}

    \begin{equation}
        \label{eq:l1_loss}
        \mathcal{L}_{L1}(G) = \mathbb{E}_{x, y} \left[ || y - G(x) ||_1 \right]
    \end{equation}
    
    \begin{equation}
        \label{eq:objective}
        G^* = \arg \min_{G} \max_{D} \mathcal{L}_{GAN}(G,D) + \lambda \mathcal{L}_{L1}(G)
    \end{equation}
    
    All trainings were run for 1000 epochs, and the best models with the lowest validation losses were saved to synthesize multichannel images in test set. We validated our model by splitting training samples using an 80:20 split. We have integrated the structural similarity index measure (SSIM) measure (Figure \ref{fig:sum}B) to detect similar spatial proteomics channels (each representing a different protein in a spatial context) to find the optimal small set of channels to condition the cGAN on such that it can synthesize a much larger set of channels that were not experimentally collected \cite{hore2010image}.
    % Figure \ref{fig:sum}C shows an output from our pipeline representing the most challenging generative synthesis task accomplished: generating a DAPI stain (a protein whose fluorescence colors nuclei, and thus captures the spatial structure of tissue) from a set of other SSIM selected channels that are not commonly associated with the spatial organization of tissue. 

\section{Generative Image Synthesis on HuBMAP Data}

      \begin{figure*}[ht!]
      \centering
      \centering
      \includegraphics[scale=0.65]{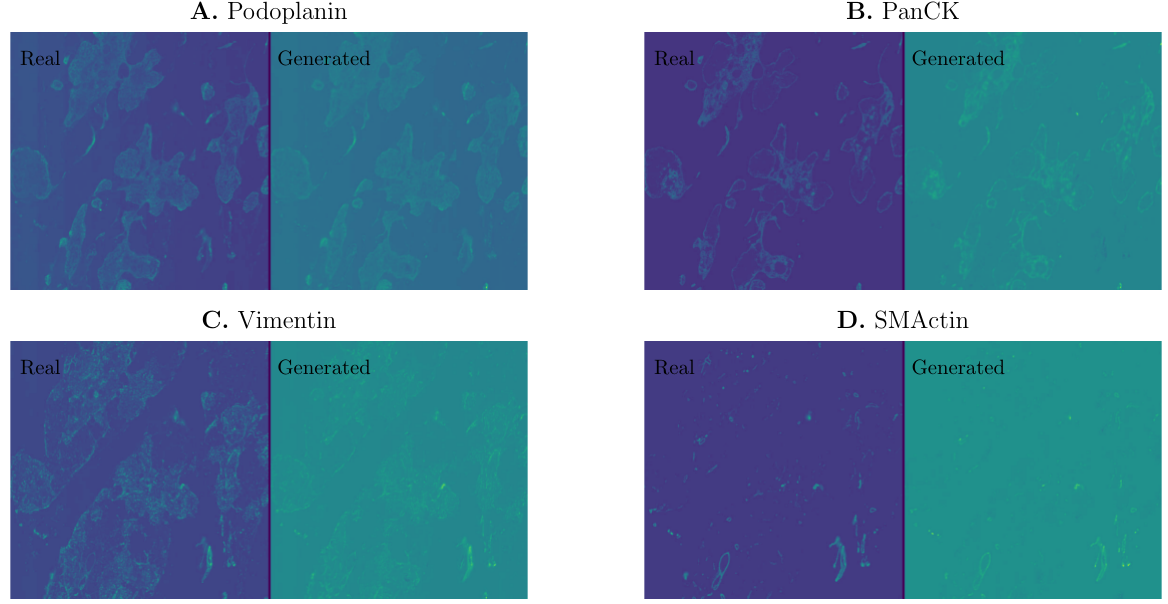}
      \caption{Real stained protein channels are displayed next to 4 generated single channel predictions. Channels shown in this figure are Podoplanin, PanCK, Vimentin, and SMActin.}
      \label{fig:hub}
      \end{figure*}
    
    As an initial pilot to test the feasibility of producing photo-accurate CODEX/PhenoCycler Fusion channels through generative synthesis, we predicted held out channels when our model was trained on publicly available images obtained through HuBMAP \cite{hubmap2019human}.
    % Figure \ref{fig:hub}A shows an interactive GIF image of the 25 training channels (see table X in Supplement). Figure \ref{fig:hub}B shows 4 held out channels (see table X2 in Supplement) that were subsequently generatively synthesized to gauge the accuracy of the algorithm, while Figure \ref{fig:hub}C shows the loss during this training. 
    For this training, 21 images with 29 protein channels each from lymph node, spleen, and thymus tissues , were downsampled to 1024 by 1024 pixels and normalized using the transform function in skimage.io \cite{van2014scikit}. The DAPI channel, which stains nuclei by binding adenine-thymine rich DNA \cite{kapuscinski1995dapi}, was near perfectly recapitulated in lymph node sections from HuBMAP training data retaining high similarity to tissue morphology and pixel intensity (Figure \ref{fig:sum}C). These results are especially surprising because the majority of channels represent stains of cell surface protein markers, highlighting the need for adequate marker selection in input training data. 

    As a result we demonstrate the ability of our model to generate photo-realistic images that retain morphological features of the underlying tissue (Figure \ref{fig:hub}).

\section{Optimal Scaling of Model and Selection of Training Channels with SSIM}

    \begin{figure*}[!ht]
      \centering
      \includegraphics[scale=0.50]{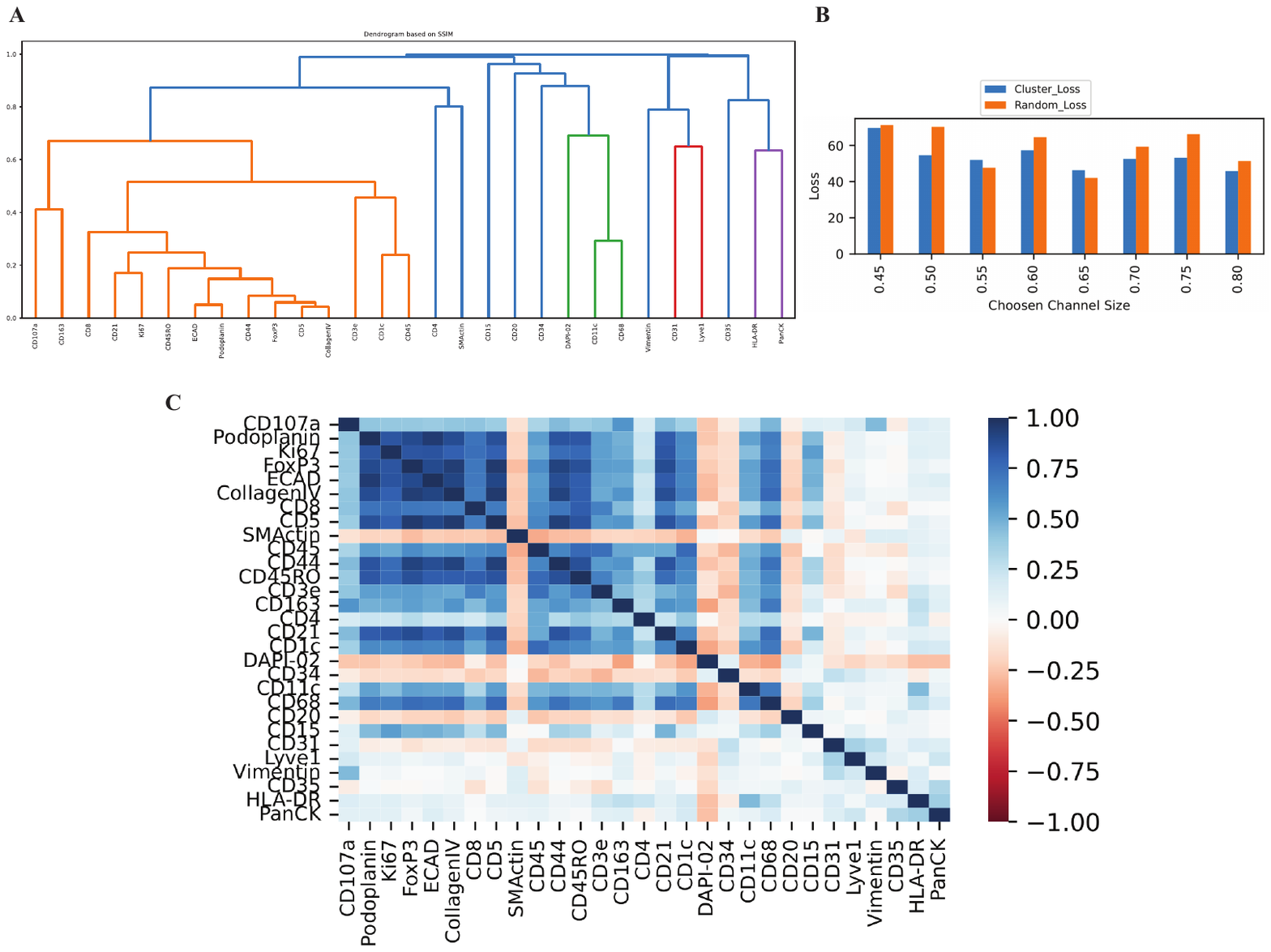}
      \caption{SSIM based clustering of multiplexed protein channels improves prediction in sparse multichannel protein data sets. (A) Clusters of single channel protein data are represented by different colors on SSIM dendrogram. (B) Loss values are reported for cluster based selection of markers in comparison to the random assignment of markers sampled from all clusters dependent upon percent of total markers sampled from clusters. (C) Correlation accuracy between single protein channels reinforces cluster assignment. Scale of correlation value indicated on right hand side shown in color value between negative one and one.}
      \label{fig:ssim}
    \end{figure*}
    
    We perform hierarchical clustering on our training channels and leverage the structural similarity index measure (SSIM) to heuristically select a minimal optimal group of channels to recapitulate the full set of channels during synthesis. For inclusion of the SSIM based channel selection in the model architecture, the derivation from Wang is used below \cite{wang2004image}. We hypothesized that individual terms comparing proteomics channels in the context of luminance (Equation \ref{eq:luminance}), contrast (Equation \ref{eq:contrast}), and structure (pixel values being affected by those in close spatial proximity, Equation \ref{eq:structure}) are each important in representing actual biological information regarding arrangement and interactions between proteins when considered in a spatial context. It is important to note that luminance, contrast, and structure can be independent and represent different aspects of underlying biology of the spatial proteome, which we have outlined below. 
    
    \begin{equation}
        \label{eq:luminance}
        l(x,y) = \frac{2\mu_x\mu_y + C_1}{\mu_x^2+ \mu_y^2 + C_1}
    \end{equation}
    
    As the CODEX/Phenocycler Fusion platform was utilized to generate all experimental data parsed in this paper, the output we have to guage spatial presence of protein in a tissue is \textit{fluorescence}, meaning that the luminance comparison term (Equation \ref{eq:luminance}) in non-simplified SSIM (Equation \ref{eq:SSIMp}) represents the actual localization and relative quantity of proteins in a channel without accounting for interactions in tissue.   
    
    \begin{equation}
        \label{eq:contrast}
        c(x,y) = \frac{2\sigma_x\sigma_y+C_2}{\sigma^2_x+\sigma^2_y+C_2}
    \end{equation}
    
    Use of a contrast comparison term (Equation \ref{eq:contrast}) as a component of SSIM is important as it indicates relative abundance of a specific protein compared against all other proteins at a specific cartesian coordinate in a slide. Contrast also represents areas of the tissue where proteins aggregate in a perturbed fashion such as complex areas at tumor/immune interfaces.
    
     \begin{equation}
        \label{eq:structure}
        s(x,y) = \frac{\sigma_{xy}+C_3}{\sigma_x\sigma_y+C_3}
    \end{equation}
    
    Use of a structure comparison term (Equation \ref{eq:structure}) contextualizes how pixels correlate with each other when they are in close spatial proximity. In the context of the spatial proteome, this can represent how complex multimers of different proteins arrange with each other spatially. 
    
    \begin{equation}
        \label{eq:SSIMp}
        \mathbb{SSIM}(x,y) =[l(x,y)]^\alpha\cdot[c(x,y)]^\beta\cdot[s(x,y)]^\gamma
    \end{equation}
    
    In the above equations, $x$ and $y$ are two images, $\alpha, \beta, \gamma$ are the weights. The pixel sample mean of x and y denoted by $\mu_x, \mu_y$. The variances of x and y are $\sigma^2_x, \sigma^2_y$. The covariance between x and y is $\sigma_{xy}$. $C_1, C_2, C_3$ are three variables used to stabilize the division.
    
    An SSIM metric including all three important yet relatively independent terms representing different facets of the underlying biology of the spatial proteome (luminance for representing position and absolute amount of protein, contrast for representing relative abundance between proteins and complex tissue structures, and structure for representing aggregation of multimers within pixel-bounded space) is displayed in Equation \ref{eq:SSIMp}. For leveraging SSIM to guide the cGAN architecture to be biologically relevant, the simplifed version from \cite{wang2004image} that is reproduced in Equation \ref{eq:SSIM} was utilized. 
    
    \begin{equation}
        \label{eq:SSIM}
        \mathbb{SSIM}(x,y) = \frac{(2\mu_x \mu_y + C_1)(2\sigma_{xy}+C_2)}{(\mu^2_x+\mu^2_y+C_1)(\sigma^2_x+\sigma^2_y+C_2)}
    \end{equation}

    This paradigm was used to hierarchically cluster protein channels from HuBMAP images (Figure \ref{fig:ssim}A). To further gauge the utility of SSIM based cluster assignments, the correlation between all pairwise protein channels in the HuBMAP data was calculated and plotted as a heatmap (Figure \ref{fig:ssim}C).
    
    We show that our channel selection algorithm minimizes training loss of the generative cGAN performing the image to image synthesis on 20\% held out test set (Figure~\ref{fig:ssim}B). This figure demonstrates that choosing training channels based on the SSIM heuristic has a considerable advantage in term of the loss reduction in almost all different scenarios of choosing channels sizes as juxtaposed to randomly selecting the same number of channels for training.

    \begin{figure*} [ht!]
      \centering
      \includegraphics[scale=0.5]{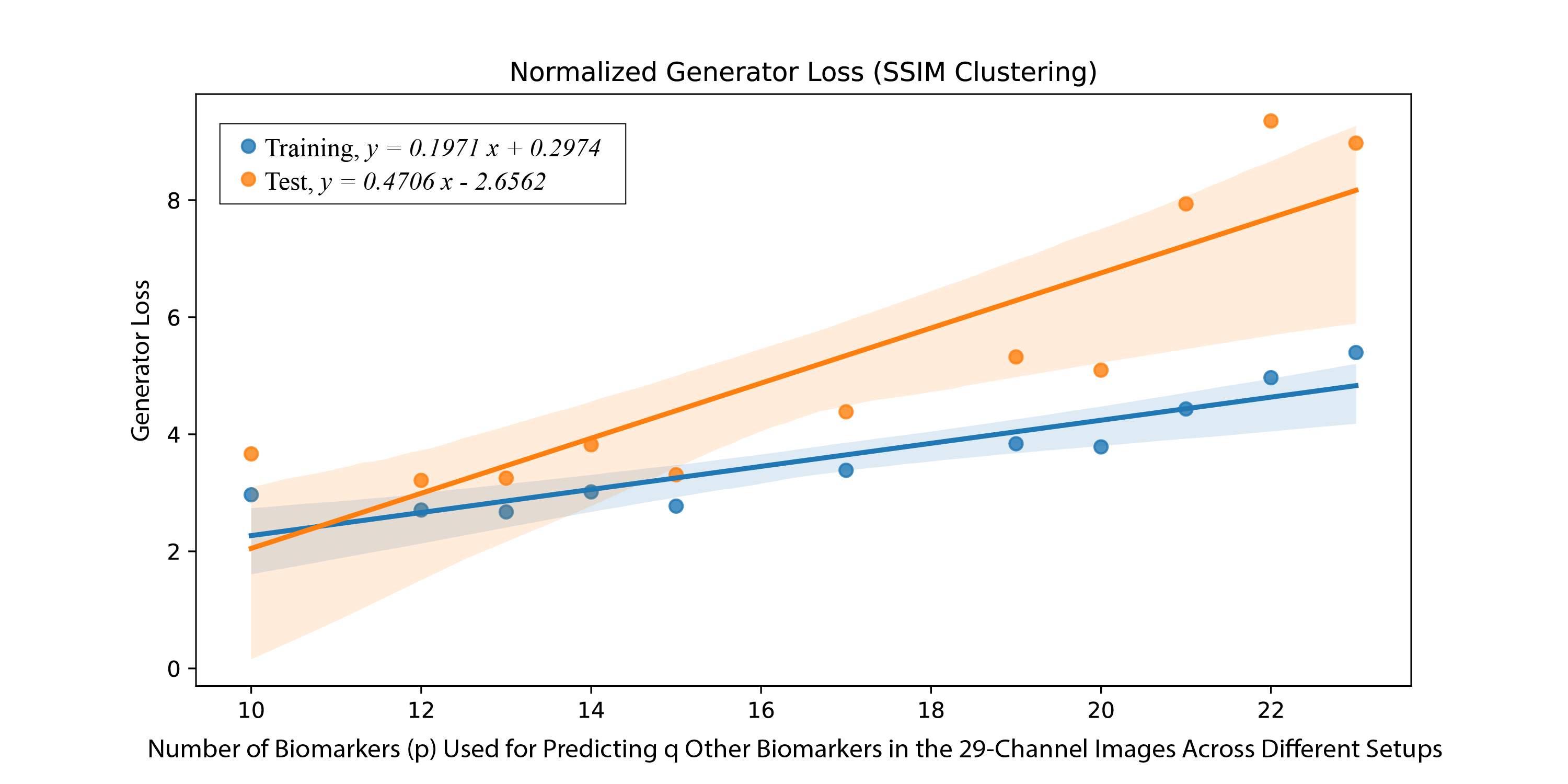}
      \caption{Normalized generator loss for training and test sets when selecting based on SSIM clustering using different number of training channels on the 29-channel HubMAP dataset.}
      \label{fig:norm_loss}
    \end{figure*}
    
     So far, we have shown how using SSIM as a heuristic leads to performance that is better than random channel selection for generative channel synthesis. Yet, with the fast pace that spatial proteomics technologies such as CODEX/PhenoCycler-Fusion are growing, we anticipate that it will be possible to multiplex hundreds of channels simultaneously in the near future. Here we introduce the concept of normalized loss as an estimate for the expected value of loss for individual prediction channels (Equation \ref{eq:norm_loss}) to discuss our model's ability to expand to images with a higher number of channels. We use this index to desensitize the loss values with respect to the number of predictive channels so that we can use the number of training channels as a proxy for the total number of channels available to train on. In simpler words, in our model settings, as the number of prediction channels increases, the number of training channels decreases making the problem twice as hard. The normalized loss takes care of the first variable, giving us a more fair index for comparing models.
     
    \begin{equation}
        \label{eq:norm_loss}
        \hat{\mathcal{L}} = \frac{\mathcal{L}}{\mbox{number of prediction channels}}
    \end{equation}
     
     Figure~\ref{fig:norm_loss} illustrates the normalized generator loss when using SSIM as a heuristic for choosing the channels to train on as the number of available channels increases. By visually inspecting our network, generated images with generator test loss of less than 65 provide informative spatial details and are considered good quality (Figure\ref{fig:ssim}B). With the same rate of expected test loss per prediction channel (regression slope = 0.4706), we can anticipate that our model would have acceptable performance for CODEX images with higher of channels. The large number of parameters in the current SSIM-guided cGAN makes it theoretically possible to fit higher dimensions.

% \section{Reconstruction of Lung Adenocarcinoma Channels from a HuBMAP Trained Model}
\section{Clinical Validation of Model}

We experimentally generated in the wet lab a new unpublished spatial proteomics dataset of lung adenocarcinoma sections using the Akoya Biosciences CODEX/Phenocycler-Fusion system. To evaluate the ability of our model to predict protein channels in generalized tissue images (i.e. from different organs and disease states than the HuBMAP model was trained on), we generated channels from new human lung adenocarcinoma tissue sections, using training data from HuBMAP samples. We took sections of lung tissue taken from 5 patients (Supplementary Table 1) and stained them for CD11c, CD15, CD21, CD31, CD4, CD8, CD45, and DAPI using the CODEX/Phenocycler-fusion platform. The per-pixel variance between real and generated DAPI channels in lung adenocarcinoma sections was comparable to that predicted for HuBMAP (Supplementary Figure 1). Predicted CD45 and CD8 channels in two lung adenocarcinoma patients showed high morphological similarity to original stains and the ability to remove artifacts from downsampling (Supplementary Figure 2). 

\section{Clinically Relevant Generative Images}
Additionally, we have tuned our generative image synthesis pipeline to create synthetic representations that we anticipate will have utility in the clinic for diagnoses. Two commonly used clinical metrics used to gauge CD8+ T Cell infiltration into solid tumors are the absolute luminance of CD8 divided by the total slide area \cite{kim2019prognostic} and reporting the ratio of the luminance of CD8 divided by the luminance of panCK, which in effect normalizes the amount of CD8 relative to tumor cells as marked by panCK \cite{monkman2020high}. 
\begin{figure*} [ht!]
      \centering
      \includegraphics[scale=0.45]{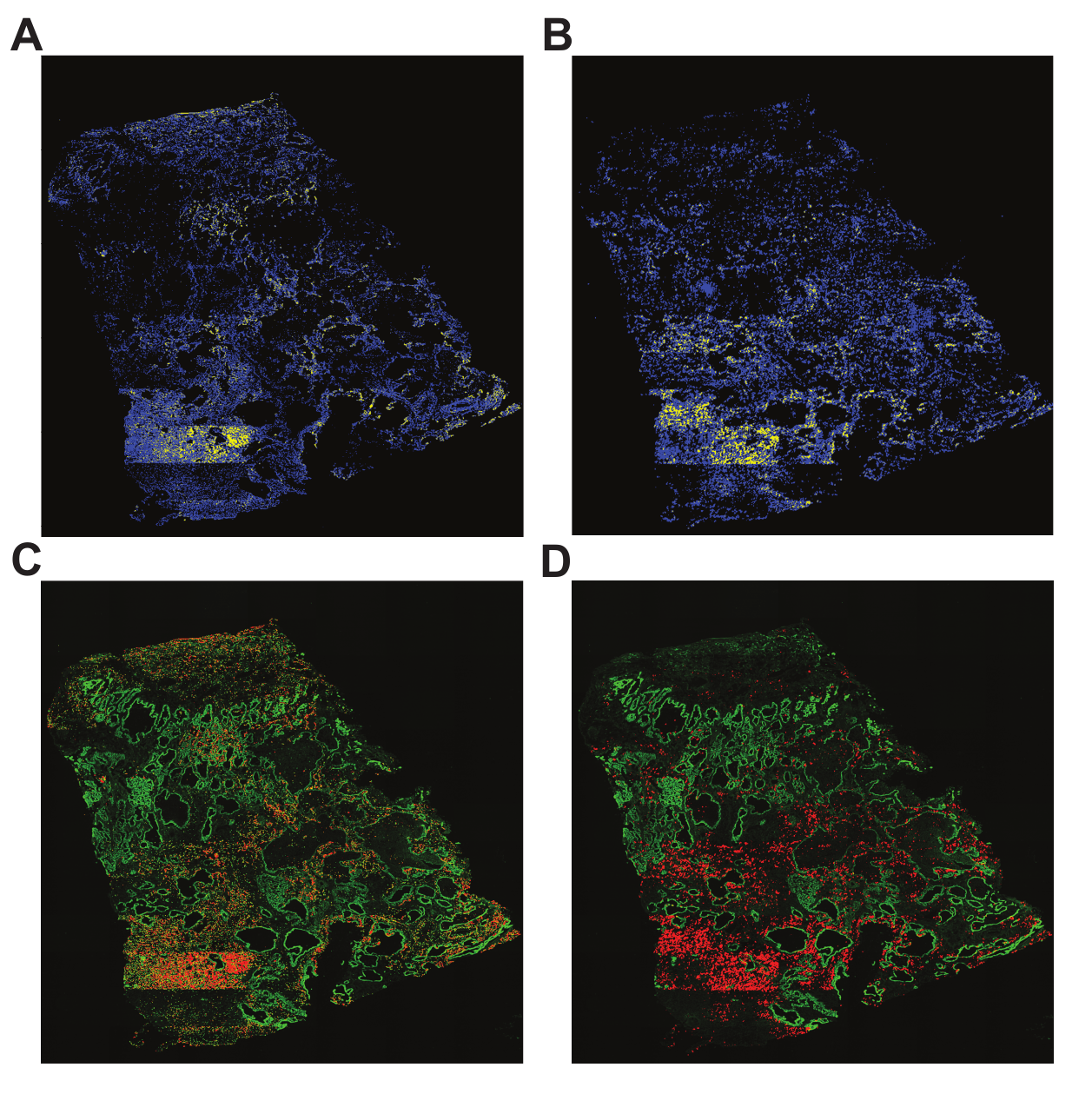}
      \caption{Generative images aiming to assist with diagnostic tasks in the clinic taken from a lung adenocarcinoma patient: A) Actual CD8 marker density (yellow) against background (blue), B) The generative version of (A) where the CD8 marker was not included as input, C) CD8 density (red) plotted with PanCK (green), D) The generative version of (C). }
      \label{fig:clinic}
    \end{figure*}
We report calculations related to these metrics in Supplementary Table 3 and show the ability to generatively synthesize visualizations of these clinical metrics in Figure \ref{fig:clinic}. 
\section{Discussion and Conclusion}
    
    This work will have important implications on the practice of medical histology and oncology, drug development at the bench, health equity, and medical ethics. The SSIM-guided cGAN algorithm we have developed is capable of recapitulating novel protein markers in large multiplexed histologies that were not experimentally included in the IHC. This method of  stain-to-stain translation has been suggested in a previous study \cite{bao2021random,bouteldja2021improving,salehi2020pix2pix,ghahremani2021deepliif} but we demonstrate that using CODEX/Phenocyler data of up to 29 multiplexed protein channels results in medically relevant loss and photo-accurate images. We confirm the ability of our cGAN to generate accurate protein stains through low pixel variance in predicted DAPI channels of HuBMAP and cancer data samples.
    
Applying stain-to-stain translation to pathological tissues presents a barrier against wide adoption of the practice because of organ and disease specific cross segmentation loss \cite{mahmood2018deep}. We address this problem by demonstrating how cluster specific markers can improve loss on multichannel predictions. Increased loss at higher cluster involvement suggests that feature bias may affect generation accuracy on specific markers.

The images that are generated from our algorithm are heavily downsampled when compared to input microscopy source files which obscures cell features in convolutional neural networks, partly due to the greater computational debt incurred when increasing generator layer dimensions \cite{salehi2020pix2pix}. Pix2pixHD solved this issue with multi-scale generative networks \cite{wang2018high}, and the implementation of visual transformers could remove the need for convolutional neural networks in the cGAN \cite{dosovitskiy2020image}. 

As our inductive proof shows, scaling this algorithm to experiments across datasets with hundreds to thousands of spatial proteomic channels will result in substantial cost savings; an example of this would be only using a small set of very expensive antibodies to generate a few SSIM optimally selected channels to gauge an effect of a new drug in the tumor microenvironment. Additionally, there will be applications related to health equity in letting hospitals in disadvantaged parts of the world synthetically generate digital pathology slides based on only a few markers.\\
One limitation of the current SSIM-guided cGAN presented in this work is that it downsamples spatial proteomics images, losing spatial information. This limitation comes from using convolutional layers. Existing methods use tile based approaches to overcome this challenge. Future iterations of this work will take the benefits of vision transformers to capture spatial information as much as possible in a full sized image.  Other areas of future exploration are considering the potential of CODEX/PhenoCycler-Fusion like methods for predicting patient survival and response; if a minimal set of markers could perform as well as the maximum possible set, this could have significant implications in decreasing the cost of certain diagnostics.\\
The ability for our cGAN to generate high quality photorealistic false images raises important ethical questions for its clinical applications. Many new guidelines will need to be crafted to ensure generative images are safely deployed in the clinic while limiting harm to patients. To start, extensive clinical trials will have to demonstrate the precision and accuracy of predictions on various patient cohorts. During this phase, patient harm may be minimized as model predictions would be benchmarked against the gold standard prior to any clinical decision-making.
However, significant injury could materialize once generative image technologies are deployed in the real-world. These errors not only pose risks to patients but could also significantly slow down the adoption of AI-based technologies, which have understandably encountered a great deal of scrutiny in healthcare. We foresee the development of quality control processes that can audit algorithms “under the hood,” akin to how modern pathology labs self-regulate their testing today. Many solutions are still being investigated, such as the manual review of outliers by pathologists; our future work will combine the SSIM based heuristic with interpretable deep Bayesian model architectures.
Lastly, as with other AI-based models, another challenge for the field of generative imaging is ensuring that all patients have equal access to these technologies and their benefits. Most datasets in digital pathology are highly skewed, resulting in models with narrow applications that may only compound pre-existing health disparities. Representation in datasets will equalize with lower infrastructure costs and higher levels of technological proficiency globally over time. In sub-Saharan Africa, significant developments in infrastructure are already underway. \cite{fatumo2022roadmap}\\
The generation of high resolution images is important in the field of computer generated data for medical purposes, as often, medical decisions are made from very small features within the image \cite{nasser2022perceptual}. Our algorithm represents the first generative model to be applied to large scale multiplexed spatial proteomic data. With the proper ethical considerations and controls, the generation of non-experimentally collected data can impact the future of diagnostic medicine and pathology for the better.

\bibliographystyle{ieeetr}
\bibliography{references.bib}

\begin{thebibliography}{10}

\bibitem{shi2011antigen}
S.-R. Shi, Y.~Shi, and C.~R. Taylor, ``Antigen retrieval immunohistochemistry:
  review and future prospects in research and diagnosis over two decades,''
  {\em Journal of Histochemistry \& Cytochemistry}, vol.~59, no.~1, pp.~13--32,
  2011.

\bibitem{tada2016potential}
H.~Tada, K.~Gonda, M.~Miyashita, and N.~Ohuchi, ``Potential clinical
  applications of next generation fluorescence immunohistochemistry for
  multiplexed and quantitative determination of biomarker in breast cancer,''
  {\em Int J Pathol Clin Res}, vol.~2, p.~022, 2016.

\bibitem{tan2020overview}
W.~C.~C. Tan, S.~N. Nerurkar, H.~Y. Cai, H.~H.~M. Ng, D.~Wu, Y.~T.~F. Wee,
  J.~C.~T. Lim, J.~Yeong, and T.~K.~H. Lim, ``Overview of multiplex
  immunohistochemistry/immunofluorescence techniques in the era of cancer
  immunotherapy,'' {\em Cancer Communications}, vol.~40, no.~4, pp.~135--153,
  2020.

\bibitem{lopes2020digital}
N.~Lopes, C.~H. Bergsland, M.~Bj{\o}rnslett, T.~Pellinen, A.~Svindland,
  A.~Nesbakken, R.~Almeida, R.~A. Lothe, L.~David, and J.~Bruun, ``Digital
  image analysis of multiplex fluorescence ihc in colorectal cancer recognizes
  the prognostic value of cdx2 and its negative correlation with sox2,'' {\em
  Laboratory Investigation}, vol.~100, no.~1, pp.~120--134, 2020.

\bibitem{stack2014multiplexed}
E.~C. Stack, C.~Wang, K.~A. Roman, and C.~C. Hoyt, ``Multiplexed
  immunohistochemistry, imaging, and quantitation: a review, with an assessment
  of tyramide signal amplification, multispectral imaging and multiplex
  analysis,'' {\em Methods}, vol.~70, no.~1, pp.~46--58, 2014.

\bibitem{chattopadhyay2012cytometry}
P.~K. Chattopadhyay and M.~Roederer, ``Cytometry: today’s technology and
  tomorrow’s horizons,'' {\em Methods}, vol.~57, no.~3, pp.~251--258, 2012.

\bibitem{goltsev2018deep}
Y.~Goltsev, N.~Samusik, J.~Kennedy-Darling, S.~Bhate, M.~Hale, G.~Vazquez,
  S.~Black, and G.~P. Nolan, ``Deep profiling of mouse splenic architecture
  with codex multiplexed imaging,'' {\em Cell}, vol.~174, no.~4, pp.~968--981,
  2018.

\bibitem{schurch2020coordinated}
C.~M. Sch{\"u}rch, S.~S. Bhate, G.~L. Barlow, D.~J. Phillips, L.~Noti,
  I.~Zlobec, P.~Chu, S.~Black, J.~Demeter, D.~R. McIlwain, {\em et~al.},
  ``Coordinated cellular neighborhoods orchestrate antitumoral immunity at the
  colorectal cancer invasive front,'' {\em Cell}, vol.~182, no.~5,
  pp.~1341--1359, 2020.

\bibitem{di2010automated}
S.~Di~Cataldo, E.~Ficarra, A.~Acquaviva, and E.~Macii, ``Automated segmentation
  of tissue images for computerized ihc analysis,'' {\em Computer methods and
  programs in biomedicine}, vol.~100, no.~1, pp.~1--15, 2010.

\bibitem{fassler2020deep}
D.~J. Fassler, S.~Abousamra, R.~Gupta, C.~Chen, M.~Zhao, D.~Paredes, S.~A.
  Batool, B.~S. Knudsen, L.~Escobar-Hoyos, K.~R. Shroyer, {\em et~al.}, ``Deep
  learning-based image analysis methods for brightfield-acquired multiplex
  immunohistochemistry images,'' {\em Diagnostic pathology}, vol.~15, no.~1,
  pp.~1--11, 2020.

\bibitem{ronneberger2015u}
O.~Ronneberger, P.~Fischer, and T.~Brox, ``U-net: Convolutional networks for
  biomedical image segmentation,'' in {\em International Conference on Medical
  image computing and computer-assisted intervention}, pp.~234--241, Springer,
  2015.

\bibitem{ghoshal2021deephistoclass}
B.~Ghoshal, F.~Hikmet, C.~Pineau, A.~Tucker, and C.~Lindskog, ``Deephistoclass:
  A novel strategy for confident classification of immunohistochemistry images
  using deep learning,'' {\em Molecular \& Cellular Proteomics}, vol.~20, 2021.

\bibitem{bulten2019epithelium}
W.~Bulten, P.~B{\'a}ndi, J.~Hoven, R.~v.~d. Loo, J.~Lotz, N.~Weiss, J.~v.~d.
  Laak, B.~v. Ginneken, C.~Hulsbergen-van~de Kaa, and G.~Litjens, ``Epithelium
  segmentation using deep learning in h\&e-stained prostate specimens with
  immunohistochemistry as reference standard,'' {\em Scientific reports},
  vol.~9, no.~1, pp.~1--10, 2019.

\bibitem{zadeh2021deep}
A.~Zadeh~Shirazi, M.~D. McDonnell, E.~Fornaciari, N.~S. Bagherian, K.~G.
  Scheer, M.~S. Samuel, M.~Yaghoobi, R.~J. Ormsby, S.~Poonnoose, D.~J. Tumes,
  {\em et~al.}, ``A deep convolutional neural network for segmentation of
  whole-slide pathology images identifies novel tumour cell-perivascular niche
  interactions that are associated with poor survival in glioblastoma,'' {\em
  British Journal of Cancer}, vol.~125, no.~3, pp.~337--350, 2021.

\bibitem{zhu2017unpaired}
J.-Y. Zhu, T.~Park, P.~Isola, and A.~A. Efros, ``Unpaired image-to-image
  translation using cycle-consistent adversarial networks,'' in {\em
  Proceedings of the IEEE international conference on computer vision},
  pp.~2223--2232, 2017.

\bibitem{tschuchnig2020generative}
M.~E. Tschuchnig, G.~J. Oostingh, and M.~Gadermayr, ``Generative adversarial
  networks in digital pathology: a survey on trends and future potential,''
  {\em Patterns}, vol.~1, no.~6, p.~100089, 2020.

\bibitem{yi2019generative}
X.~Yi, E.~Walia, and P.~Babyn, ``Generative adversarial network in medical
  imaging: A review,'' {\em Medical image analysis}, vol.~58, p.~101552, 2019.

\bibitem{yi2018sharpness}
X.~Yi and P.~Babyn, ``Sharpness-aware low-dose ct denoising using conditional
  generative adversarial network,'' {\em Journal of digital imaging}, vol.~31,
  no.~5, pp.~655--669, 2018.

\bibitem{gadermayr2018way}
M.~Gadermayr, V.~Appel, B.~M. Klinkhammer, P.~Boor, and D.~Merhof, ``Which way
  round? a study on the performance of stain-translation for segmenting
  arbitrarily dyed histological images,'' in {\em International Conference on
  Medical Image Computing and Computer-Assisted Intervention}, pp.~165--173,
  Springer, 2018.

\bibitem{gupta2019gan}
L.~Gupta, B.~M. Klinkhammer, P.~Boor, D.~Merhof, and M.~Gadermayr, ``Gan-based
  image enrichment in digital pathology boosts segmentation accuracy,'' in {\em
  International Conference on Medical Image Computing and Computer-Assisted
  Intervention}, pp.~631--639, Springer, 2019.

\bibitem{bao2021random}
Y.~Bao, S.~amd Tang, H.~H. Lee, R.~Gao, S.~Chiron, I.~Lyu, L.~A. Coburn, K.~T.
  Wilson, J.~Roland, B.~A. Landman, {\em et~al.}, ``Random multi-channel image
  synthesis for multiplexed immunofluorescence imaging,'' in {\em MICCAI
  Workshop on Computational Pathology}, pp.~36--46, PMLR, 2021.

\bibitem{schapiro2022mcmicro}
D.~Schapiro, A.~Sokolov, C.~Yapp, Y.-A. Chen, J.~L. Muhlich, J.~Hess, A.~L.
  Creason, A.~J. Nirmal, G.~J. Baker, M.~K. Nariya, {\em et~al.}, ``Mcmicro: A
  scalable, modular image-processing pipeline for multiplexed tissue imaging,''
  {\em Nature methods}, vol.~19, no.~3, pp.~311--315, 2022.

\bibitem{schapiro2022miti}
D.~Schapiro, C.~Yapp, A.~Sokolov, S.~M. Reynolds, Y.-A. Chen, D.~Sudar, Y.~Xie,
  J.~Muhlich, R.~Arias-Camison, S.~Arena, {\em et~al.}, ``Miti minimum
  information guidelines for highly multiplexed tissue images,'' {\em Nature
  methods}, vol.~19, no.~3, pp.~262--267, 2022.

\bibitem{Gavrikov2020VisualKeras}
P.~Gavrikov, ``visualkeras.''
  \url{https://github.com/paulgavrikov/visualkeras}, 2020.

\bibitem{https://doi.org/10.48550/arxiv.1711.11585}
T.-C. Wang, M.-Y. Liu, J.-Y. Zhu, A.~Tao, J.~Kautz, and B.~Catanzaro,
  ``High-resolution image synthesis and semantic manipulation with conditional
  gans,'' 2017.

\bibitem{jeong2020solar}
H.-J. Jeong, Y.-J. Moon, E.~Park, and H.~Lee, ``Solar coronal magnetic field
  extrapolation from synchronic data with ai-generated farside,'' {\em The
  Astrophysical Journal Letters}, vol.~903, no.~2, p.~L25, 2020.

\bibitem{isola2017image}
P.~Isola, J.-Y. Zhu, T.~Zhou, and A.~A. Efros, ``Image-to-image translation
  with conditional adversarial networks,'' in {\em Proceedings of the IEEE
  conference on computer vision and pattern recognition}, pp.~1125--1134, 2017.

\bibitem{hore2010image}
A.~Hore and D.~Ziou, ``Image quality metrics: Psnr vs. ssim,'' in {\em 2010
  20th international conference on pattern recognition}, pp.~2366--2369, IEEE,
  2010.

\bibitem{hubmap2019human}
H.~Consortium, ``The human body at cellular resolution: the nih human
  biomolecular atlas program,'' {\em Nature}, vol.~574, no.~7777, p.~187, 2019.

\bibitem{van2014scikit}
S.~Van~der Walt, J.~L. Sch{\"o}nberger, J.~Nunez-Iglesias, F.~Boulogne, J.~D.
  Warner, N.~Yager, E.~Gouillart, and T.~Yu, ``scikit-image: image processing
  in python,'' {\em PeerJ}, vol.~2, p.~e453, 2014.

\bibitem{kapuscinski1995dapi}
J.~Kapuscinski, ``Dapi: a dna-specific fluorescent probe,'' {\em Biotechnic \&
  histochemistry}, vol.~70, no.~5, pp.~220--233, 1995.

\bibitem{wang2004image}
Z.~Wang, A.~C. Bovik, H.~R. Sheikh, and E.~P. Simoncelli, ``Image quality
  assessment: from error visibility to structural similarity,'' {\em IEEE
  transactions on image processing}, vol.~13, no.~4, pp.~600--612, 2004.

\bibitem{kim2019prognostic}
S.-H. Kim, S.-I. Go, D.~H. Song, S.~W. Park, H.~R. Kim, I.~Jang, J.~D. Kim,
  J.~S. Lee, and G.-W. Lee, ``Prognostic impact of cd8 and programmed
  death-ligand 1 expression in patients with resectable non-small cell lung
  cancer,'' {\em British journal of cancer}, vol.~120, no.~5, pp.~547--554,
  2019.

\bibitem{monkman2020high}
J.~Monkman, T.~Taheri, M.~Ebrahimi~Warkiani, C.~O’leary, R.~Ladwa,
  D.~Richard, K.~O’Byrne, and A.~Kulasinghe, ``High-plex and high-throughput
  digital spatial profiling of non-small-cell lung cancer (nsclc),'' {\em
  Cancers}, vol.~12, no.~12, p.~3551, 2020.

\bibitem{bouteldja2021improving}
N.~Bouteldja, B.~M. Klinkhammer, T.~Schlaich, P.~Boor, and D.~Merhof,
  ``Improving unsupervised stain-to-stain translation using self-supervision
  and meta-learning,'' {\em arXiv preprint arXiv:2112.08837}, 2021.

\bibitem{salehi2020pix2pix}
P.~Salehi and A.~Chalechale, ``Pix2pix-based stain-to-stain translation: a
  solution for robust stain normalization in histopathology images analysis,''
  in {\em 2020 International Conference on Machine Vision and Image Processing
  (MVIP)}, pp.~1--7, IEEE, 2020.

\bibitem{ghahremani2021deepliif}
P.~Ghahremani, Y.~Li, A.~Kaufman, R.~Vanguri, N.~Greenwald, M.~Angelo, T.~J.
  Hollmann, and S.~Nadeem, ``Deepliif: Deep learning-inferred multiplex
  immunofluorescence for ihc image quantification,'' {\em bioRxiv}, 2021.

\bibitem{mahmood2018deep}
F.~Mahmood, D.~Borders, R.~Chen, G.~McKay, K.~Salimian, A.~Baras, and N.~Durr,
  ``Deep adversarial training for multi-organ nuclei segmentation in
  histopathology images. arxiv preprint,'' {\em arXiv preprint
  arXiv:1810.00236}, 2018.

\bibitem{wang2018high}
T.-C. Wang, M.-Y. Liu, J.-Y. Zhu, A.~Tao, J.~Kautz, and B.~Catanzaro,
  ``High-resolution image synthesis and semantic manipulation with conditional
  gans,'' in {\em Proceedings of the IEEE conference on computer vision and
  pattern recognition}, pp.~8798--8807, 2018.

\bibitem{dosovitskiy2020image}
A.~Dosovitskiy, L.~Beyer, A.~Kolesnikov, D.~Weissenborn, X.~Zhai,
  T.~Unterthiner, M.~Dehghani, M.~Minderer, G.~Heigold, S.~Gelly, {\em et~al.},
  ``An image is worth 16x16 words: Transformers for image recognition at
  scale,'' {\em arXiv preprint arXiv:2010.11929}, 2020.

\bibitem{fatumo2022roadmap}
S.~Fatumo, T.~Chikowore, A.~Choudhury, M.~Ayub, A.~R. Martin, and
  K.~Kuchenbaecker, ``A roadmap to increase diversity in genomic studies,''
  {\em Nature Medicine}, vol.~28, no.~2, pp.~243--250, 2022.

\bibitem{nasser2022perceptual}
S.~A. Nasser, S.~Shamsi, V.~Bundele, B.~Garg, and A.~Sethi, ``Perceptual cgan
  for mri super-resolution,'' {\em arXiv preprint arXiv:2201.09314}, 2022.

\end{thebibliography}

\end{document}